\documentclass[a4paper,journal]{IEEEtran}

\IEEEoverridecommandlockouts
\usepackage{amssymb,amsmath}
\usepackage{graphicx}
\usepackage{epstopdf}
\usepackage{accents}
\usepackage{pgfplots}
\usepackage{pgfplotstable}
\DeclareGraphicsExtensions{.eps,.pdf} 
\graphicspath{ {figures/} }

\usepackage{authblk}
\usepackage{geometry}
\usepackage{color}
\usepackage{array}
\usepackage{cite}
\usepackage{subfigure}
\usepackage{amssymb,amsmath}
\usepackage{algorithmic}
\usepackage{color}
\usepackage{epstopdf}
\usepackage{multirow}
\usepackage{multicol}
\usepackage{cases}
\usepackage{setspace}
\usepackage{caption}
\usepackage{mathrsfs}
\usepackage{delarray}
\usepackage{tikz}
\usepackage[ruled]{algorithm2e}
\usepackage{balance}
\usepackage[mathscr]{euscript}
 \let\mathscr\relax
\usepackage[scr]{rsfso}

\allowdisplaybreaks
\usepackage{bbm}

\usepackage[mathscr]{euscript}
 \let\mathscr\relax
\usepackage[scr]{rsfso}

\newtheorem{remark}{{\textbf{Remark}}}

\newcommand{\st}{{\mathrm{s.t.}}}

\renewcommand{\algorithmicrequire}{\textbf{Input:}}

     \usepackage[compact]{titlesec}
    \titlespacing{\section}{0pt}{2ex}{1ex}
    \titlespacing{\subsection}{0pt}{1ex}{0ex}
    \titlespacing{\subsubsection}{0pt}{0.5ex}{0ex}

\newcommand{\sfT}{\mathsf{T}}

\usepackage{capt-of}
\usepackage{dblfloatfix}
\usepackage{varwidth}
\usepackage{tikz}

\usepackage[mathscr]{euscript}
 \let\mathscr\relax
\usepackage{bbm}


\usepackage{babel}
\usepackage{algorithm2e}

\begin{document}
\title{Energy-Aware Resource Allocation for Energy Harvesting Powered Wireless Sensor Nodes}
\author{Ngoc M. Ngo, Trung T. Nguyen, Phuc H. Nguyen and Van-Dinh Nguyen, \textit{Senior Member, IEEE}
\thanks{The authors are with the College of Engineering and Computer Science, VinUniversity, Vinhomes Ocean Park, Hanoi 100000, Vietnam (e-mail: dinh.nv2@vinuni.edu.vn). This work is sponsored by VinUniversity under Grant No. VUNI.GT.NTKH.06.}
}

\maketitle

\begin{abstract}
Low harvested energy poses a significant challenge to sustaining continuous communication in energy harvesting (EH)-powered wireless sensor networks. This is mainly due to intermittent and limited power availability from radio frequency signals. In this paper, we introduce a novel energy-aware resource allocation problem aimed at enabling the asynchronous accumulate-then-transmit protocol, offering an alternative to the extensively studied harvest-then-transmit approach. Specifically, we jointly optimize power allocation and time fraction dedicated to EH to maximize the average long-term system throughput, accounting for both data and energy queue lengths. By leveraging inner approximation and network utility maximization techniques, we develop a simple yet efficient iterative algorithm that guarantees at least a local optimum and achieves long-term utility improvement. Numerical results highlight the proposed approach's effectiveness in terms of both queue length and sustained system throughput.
\end{abstract}

\section{Introduction}\label{sec_Intro}
\IEEEPARstart{W}{ireless} sensor networks (WSNs) have gained widespread attention due to their potential applications in environmental monitoring, smart agriculture, and healthcare. However, the limited energy resources of battery-powered nodes pose challenges to network longevity and reliability. Traditional solutions often rely on regular battery replacement, which is impractical in remote or inaccessible environments. This has led to a growing interest in energy harvesting (EH) technologies as a sustainable alternative to power wireless sensor nodes by harnessing ambient energy sources, such as solar, vibration, or radio frequency (RF) signals \cite{KaswanCOMST22}.

To achieve a manageable solution, advancing wireless energy transfer (WET) technology offers a promising solution. WET facilitates downlink energy harvesting followed by uplink wireless information transmission (WIT) \cite{MakTCOM16,ChenTWC19}. This approach has been investigated in various network scenarios, demonstrating its potential to support mission-critical IoT applications \cite{WangTWC19} and improve sum throughput in IoT networks \cite{NguyenTCOM17}. For instance, the authors in \cite{CheCOMLL15} proposed an optimal closed-form solution that jointly optimizes energy and information beamforming, effectively balancing energy transfer efficiency with co-channel interference. Additionally, the work in \cite{NguyenTWC23} examined the application of WET in multi-hop networks, focusing on achieving high block error rate (BLER) performance alongside reliable and low-latency communications.

Despite the promise of EH techniques, sustaining continuous communication in EH-powered WSNs remains particularly challenging. Previous works have predominantly relied on the harvest-then-transmit protocol \cite{WangTWC19,NguyenTCOM17,CheCOMLL15,NguyenTWC23, ChenTWC19}, where harvested energy is entirely consumed within a single time slot. This approach inherently limits the ability to support sustained communication over time, making it unsuitable for scenarios requiring long-term reliability. The primary challenge arises from the intermittent and limited power supply of ambient energy sources, particularly RF signals \cite{KaswanCOMST22,MakTCOM16}, which are inherently unpredictable and often insufficient. These energy fluctuations disrupt data transmission and significantly degrade network performance, underscoring the critical need for advanced resource allocation techniques that can dynamically balance energy harvesting and data transmission to maintain long-term network stability and efficiency.

This paper addresses the energy-aware resource allocation problem in EH-powered WSNs. The asynchronous accumulate-then-transmit protocol allows nodes to store harvested energy over multiple time slots before transmission, unlike the conventional harvest-then-transmit protocol. This flexibility enables improved power allocation and adaptation to fluctuating energy and data traffic. Our approach aims to maximize long-term system throughput while managing the dynamic nature of data and energy queues. We propose an iterative optimization algorithm utilizing inner approximation (IA) \cite{Beck:JGO:10} and network utility maximization (NUM) techniques \cite{neely2010stochastic}, which ensures convergence to at least a local optimum and achieves improved long-term utility performance. Numerical simulations validate the effectiveness of our approach, demonstrating significant improvements in overall system throughput while achieving a better tradeoff between throughput and stability compared to existing methods.

\section{System Model and Problem Formulation}\label{sec_SystemModel}
\subsection{System Model}

 \begin{figure}[h]
 	\centering
 	\includegraphics[width=.9\columnwidth,trim={0cm 0.0cm 0cm 0.0cm}]{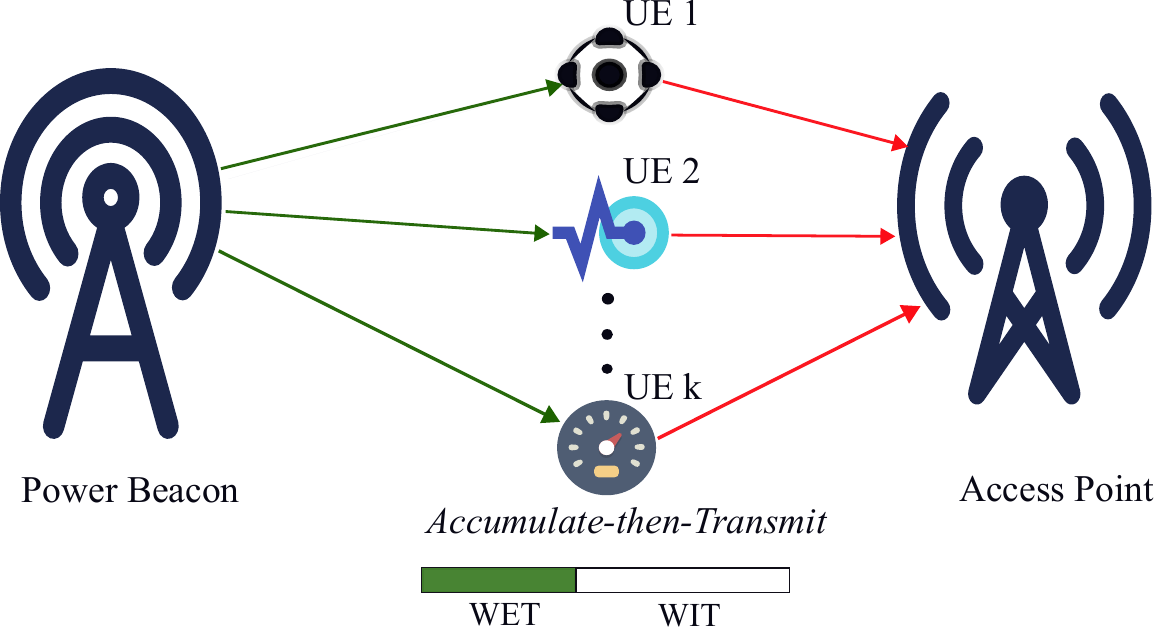}
 	\caption{\small The aynchronous accumulate-then-transmit-based system model.}
 	\label{fig:Systemmodel}
 \end{figure}

Fig. \ref{fig:Systemmodel} depicts an EH-powered wireless sensor network, which includes a power beacon (PB) and an access point (AP) serving a set $\mathcal{K}\triangleq\{1,\cdots, K\}$ of $K$ single-antenna sensors (\textit{i.e.} called user equipment (UE) for convenience). The PB and AP are equipped with $M$ and $N$ antennas, respectively. 
The system operates in discrete time, indexed by $t \in \mathcal{T}\triangleq \{1, 2, \dots, T\}$. Let $\tau$ represent the duration of each time-frame $t$. The channels from PB to UE $k$ and from UE $k$ to AP at time $t$ are denoted by $\boldsymbol{g}_k[t] \in \mathbb{C}^M$ and $\boldsymbol{h}_k[t] \in \mathbb{C}^N$, respectively. We assume a low-mobility scenario where all wireless links undergo quasi-static flat fading, experiencing both large- and small-scale fading, and remain constant during each time-frame $t$.

We employ frequency-division multiple access (FDMA) for both WET and WIT links, utilizing a total system bandwidth of $W$. Each user is allocated an equal bandwidth of $w_k = W/K$ to harvest energy and transmit information to AP. During each time-frame $t$, each UE uses its dedicated bandwidth for both WET and WIT processes, as illustrated in Fig. \ref{fig:Systemmodel}. Specifically, the time fraction allocated for WET by UE $k$ in time-frame $t$ is denoted as $\alpha_k[t] \in [0, 1]$, while the remaining time fraction, $(1 - \alpha_k[t])$, is used for WIT.

\noindent\textit{1) WET Process}:
Let $P_{\max}$ be the total power budget of PB. Under the FDMA and linear EH model \cite{MakTCOM16,ChenTWC19}, the energy harvested by UE $k$ during time-frame $t$ can be expressed as
\begin{equation} \label{eq1}
e_k[t] = \eta\tau\alpha_k[t] p_k^e[t]\|\boldsymbol{g}_k[t]\|^2,\, \forall  k\in\mathcal{K}
\end{equation}
where $\eta\in(0,1)$ denotes the energy conversion efficiency, and $p_k^e[t]$ is the power allocated to UE $k$ by PB, such as \textit{i.e.} $\sum_{k\in\mathcal{K}}p_k^e[t] \leq P_{\max}$. We note that $e_k[t]$ is unknown to PB until the end of the WET phase. Given  that $p_k^e[t]\leq P_{\max}$ and $\alpha_k[t]\leq 1$, it follows that $0 \leq e_k[t] \leq \eta\tau P_{\max} M$.

\noindent\textit{2) WIT Process}:
During the remaining time fraction,  we utilize the maximum ratio combining (MRC) technique to convey messages from UEs to AP in an asynchronous fashion. The  signal received at AP for the intended signal of UE $k$ at time-frame $t$ can be expressed as
\begin{equation}
\boldsymbol{y}_k[t] =  \sqrt{p_k^i[t]}\boldsymbol{h}_k[t]x_k[t] + \boldsymbol{n}_k[t],\, \forall  k\in\mathcal{K}
\end{equation}
where $p_k^i[t]$ and $x_k[t]$ with $\mathbb{E}\{|x_k[t]|^2\}=1$ are the transmission power and transmitted signal of UE $k$, respectively, while  $\boldsymbol{n}_k[t]\sim\mathcal{CN}(0,\sigma^2\mathbf{I}_N)$ is the AWGN. As a result, the signal-to-noise ratio (SNR) of UE $k$ at time-frame $t$ is given by
\begin{equation}
\gamma_k[t] =  \frac{p_k^i[t]\|\boldsymbol{h}_k[t]\|^2}{\sigma^2},\, \forall  k\in\mathcal{K}.
\end{equation}
It follows that $\gamma_k[t]$ is bounded by $\gamma_k[t] \leq \bar{P}_k N/\sigma^2, \forall k$, where $\bar{P}_k$ is the maximum power transmission allowed for UE $k$ due to hardware limitation. By $\boldsymbol{p}[t]\triangleq\{p_k^i[t],p_k^e[t]\}_{\forall k}$, the achievable throughput of UE $k$ within the WIT phase is expressed as
\begin{equation}
r_k(p_k^i[t]) =  w_k\log_2\big(1 +\gamma_k[t] \big), \, \forall  k\in\mathcal{K}.
\end{equation}

\noindent\textit{3) Queueing Models}

\noindent\textbf{Data queue model}: Each UE is assumed to maintain an individual data queue. The data arrival rate at UE $k$ at time-frame $t$, denoted by $a_k[t]$ (bits/s), has a mean value of $\mathbb{E}\{a_{k}[t]\}=\bar{a}_k$. We assume that $a_k[t]$ is randomly generated and upper bounded by a finite constant $a^{\max}$, such as $a_k[t] \leq a^{\max}<\infty,\, \forall k$. The queue-length of UE $k$  evolves as follows: 
\begin{equation}\label{eq:DATAqueue}
q_k[t+1] = \bigl[q_k[t] + a_k[t]\tau - r_k(p_k^i[t])(1-\alpha_k[t])\tau \bigr]^+ 
\end{equation}
where $[x]^+\triangleq\max\{0,x\}$. By $\boldsymbol{q}[t]\triangleq\bigl[q_k[t] \bigr]^{\sfT}_{\forall k}$, we consider a queueing network as \textit{stable} if the steady-state total queue length remains finite \cite{EryilmazJSAC2006}:
\begin{align}\label{steady-state}
    \underset{t\to\infty}{\limsup}\ \mathbb{E}\{\|\boldsymbol{q}[t]\|_1\} < \infty.
\end{align}

\noindent\textbf{Energy queue model}: We denote by $E_k^{\max}$ the maximum battery capacity of UE $k$. By the accumulate-then-transmit protocol \cite{AkinCOMML17} and \eqref{eq:DATAqueue}, the energy queue backlog of UE $k$, denoted by $E_k[t]$, evolves as follows:
\begin{align}\label{eq:EHqueue_1}
    E_k[t+1] = &\min\big\{E_k[t] + e_k[t] \nonumber\\
    &\qquad-  p_k^i[t](1-\alpha_k[t])\tau,\, E_k^{\max} \big\},\, \forall  k.
\end{align}

\subsection{Problem Formulation}
At the beginning of each time-frame $t$, the concrete information about the system state is defined as $\boldsymbol{s}[t]\triangleq(s_k[t], \forall k)\triangleq(e_k[t], a_k[t], \forall k)$. In this work, our goal is to maximize the average long-term network utility function $U(\boldsymbol{p}|\boldsymbol{s})$ over $T$ time-frames:
\begin{align}\label{eq:EHqueue}
U(\boldsymbol{p}|\boldsymbol{s}) & = \sum\nolimits_{k\in\mathcal{K}}\mathbb{E}\{r_k(p_k^i[t]) \}\nonumber\\
&\triangleq \sum\nolimits_{k\in\mathcal{K}}\mathbb{E}\{w_k\log_2\big(1 +\gamma_k[t] \big)\}.
\end{align}
The effective energy available at UE $k$ in the WIT phase can be calculated as
\begin{equation} \label{eq9}
e^{\mathtt{eff}}_k[t] = E_k[t] + e_k[t],\, \forall k, t.
\end{equation}

Summing up, the energy-aware resource allocation problem can be mathematically formulated as
\begin{subequations} \label{eq:JFCS1}
	\begin{IEEEeqnarray}{cl}
		 \underset{\boldsymbol{p}, \boldsymbol{\alpha}}{\mathrm{max}}& \   U(\boldsymbol{p}|\boldsymbol{s})\label{eq:problema} \\
		 \st &  \  \underset{t\to\infty}{\limsup}\ \mathbb{E}\{\|\boldsymbol{q}[t]\|_1\} < \infty                       \label{eq:problemab}\\
		     & \  \sum\nolimits_{k\in\mathcal{K}}p_k^e[t] \leq P_{\max},\,\forall t                      \label{eq:problemac}\\
		    &\ p_k^i[t] \leq \min\Big\{\frac{e^{\mathtt{eff}}_k}{(1-\alpha_k[t])\tau}, \bar{P}_k\Big\},\, \forall  k, t\label{eq:problemad} \qquad\\
		     &\ p_k^i[t] \geq \underline{p}^{\min}, \alpha_k[t]\in[0,1],\, \forall  k, t \label{eq:problemaf} \quad
 \end{IEEEeqnarray}
\end{subequations}
where $\boldsymbol{\alpha}[t] = \{\alpha_k[t]\}_{\forall k}$, and $\underline{p}^{\min} \triangleq \frac{\underline{\gamma}^{\min}\sigma^2}{\|\boldsymbol{h}_k[t]\|^2}$ with $\underline{\gamma}^{\min}$ being the minimum SNR requirement. In \eqref{eq:problemad}, we consider that the transmission power of each UE $k$ in each time-frame $t$ is restricted by the available harvested power $e^{\mathtt{eff}}_k/(1-\alpha_k[t])\tau$ and $\bar{P}_k$ due to hardware limitation.
\begin{remark}
 The achievable throughput $r_k(p_k^i[t])$ is twice continuously differentiable, increasing, and concave in $p_k^i[t]$ (\textit{i.e.} $r_k^{''}(p_k^i[t]) <0$). In addition, it follows that $r_k(p_k^i[t])\leq W\log_2(1+\frac{\bar{P}_k N}{\sigma^2})< \infty$.
\end{remark}

\section{Proposed Solution}
\textit{Challenge of solving problem \eqref{eq:JFCS1}}: 
The non-convex nature of constraint \eqref{eq:problemad} makes it difficult to solve problem \eqref{eq:JFCS1} optimally. Furthermore, the expectations in \eqref{eq:problema} and \eqref{eq:problemab} introduce stochastic elements, which cannot be addressed directly using the IA framework \cite{Beck:JGO:10}.

\subsection{The Overall Algorithm}
To address the challenge of handling \eqref{eq:problemab}, we define the total queue backlog of all UEs in time-slot $t$ as $L\left[t\right] = \frac{1}{2}\sum_{k \in \mathcal{K}} \frac{{q_k}\left[t\right]^2}{\tau^2}$, which is represented as a quadratic Lyapunov function \cite{neely2010stochastic,nguyen_2023_networkaided}. The Lyapunov drift from time-slot $t$ to $t+1$ is given as $\Delta L[t] \triangleq L[t+1] - L[t]$.  To bypass the direct handling of constraint \eqref{eq:problemab}, we penalize the Lyapunov drift, \textit{i.e.} $\underset{\boldsymbol{p}, \boldsymbol{\alpha}}{\mathrm{max}}\,U(\boldsymbol{p}|\boldsymbol{s}) - \mathbb{E}\left\{\Delta L\left[t\right]\right\}$. By the drift-plus-penalty procedure \cite{neely2010stochastic}, we rewrite \eqref{eq:JFCS1} as the following penalized optimization problem: 
\begin{subequations} \label{eq:JFCS2}
    \begin{IEEEeqnarray}{cl}
		\, & \underset{\boldsymbol{p}, \boldsymbol{\alpha}}{\mathrm{max}} \ \beta U(\boldsymbol{p}|\boldsymbol{s}) -\, \mathbb{E}\left\{\Delta L\left[t\right]\right\}    \label{eq:problemg} \\
        & \st\quad \eqref{eq:problemac}, \eqref{eq:problemad}, \eqref{eq:problemaf}
    \end{IEEEeqnarray}
\end{subequations}
where $\beta$ is a positive scaling factor to guarantee network stability (\textit{i.e.} \eqref{eq:problemab}). When $\beta$ is small, problem \eqref{eq:JFCS2} places greater emphasis on minimizing $\Delta L$, prioritizing queue stability over throughput, and vice versa. To bound the Lyapunov drift, we introduce slack variables
 $\boldsymbol{\lambda[t]}\triangleq \big\{\lambda_k[t]\big\}_{\forall k}$, leading to $\Delta L\left[t\right]$ as $\Delta L^{\mathtt{UB}}[t] \triangleq \frac{1}{2} \sum_{k \in \mathcal{K}} \frac{\lambda_k^2[t]}{\tau^2} - L[t] \geq \Delta L[t]$, which imposes the following constraint:
\begin{equation} \label{eq:12}
    q_k[t+1] \leq \lambda_k[t],\,\forall k.
\end{equation}
It is clear that $\Delta L^{\mathtt{UB}}[t] = \Delta L[t]$ if the inequality \eqref{eq:12} holds. Since $L[t]$ is known at time-slot $t$, the upper bounded
Lyapunov drift function is convex in $(\boldsymbol{p}, \boldsymbol{\alpha})$. To facilitate the optimization process, we solve \eqref{eq:JFCS1} in each time-slot $t$ as:
\begin{subequations} \label{eq:JFCS2_2}
    \begin{IEEEeqnarray}{cl}
		\underset{\boldsymbol{p}[t], \boldsymbol{\alpha}[t], \boldsymbol{\lambda}[t]}{\mathrm{max}}\,&  \beta \sum_{k\in\mathcal{K}}r_k(p_k^i[t]) - \Delta L^{\mathtt{UB}}[t]
        \label{eq:JFCS2_2:a} \\
        \st&   \eqref{eq:problemac}, \eqref{eq:problemad}, \eqref{eq:problemaf}, \eqref{eq:12} \label{eq:JFCS2_2:b}.
    \end{IEEEeqnarray}
\end{subequations}
The overall proposed algorithm for solving problem \eqref{eq:JFCS1} is summarized in Algorithm \ref{Alg1}. In each time-frame, we solve problem \eqref{eq:JFCS2_2} by Algorithm \ref{alg2} presented next and subsequently update queue lengths.

\begin{algorithm}[t]
	\begin{algorithmic}[1]
 		\fontsize{9}{9}\selectfont
		\protect\caption{Overall Algorithm for Solving  Problem \eqref{eq:JFCS1}}\label{Alg1}
       \global\long\def\algorithmicrequire{\textbf{Initialization:}}	
      \REQUIRE Set $t=1$ and choose a positive scaling factor $\beta$. All queues are set to be empty: $q_k[1]=0$ and $E_k[1]=0, \forall k\in\mathcal{K}$.
   
		\FOR{each time-frame $t\in\{1,2,\cdots,T\}$}
	   \STATE \textbf{Call Algorithm \ref{alg2}} to solve problem \eqref{eq:JFCS2_2}: Obtain the optimal solution $\{\boldsymbol{p}^{(*)}[t], \boldsymbol{\alpha}^{(*)}[t], \boldsymbol{\lambda}^{(*)}[t]\}$;
	    
	\STATE Queue-length updates: 
	       \begin{align}
	          q_k[t+1] = &\bigl[q_k[t] + a_k[t]\tau \nonumber\\
            &- r_k(p_k^{i,(*)}[t])(1-\alpha_k^{(*)}[t])\tau \bigr]^+, \forall  k\qquad\nonumber\\
	         E_k[t+1] = &\min\big\{E_k[t] + e_k[t]-  p_k^{i,(*)}[t]\times \nonumber\\
    &\qquad(1-\alpha_k^{(*)}[t])\tau,\, E_k^{\max} \big\},\, \forall  k.\nonumber
	       \end{align}
	
           \STATE Set $t=t+1$;
        \ENDFOR
\end{algorithmic}
\end{algorithm}

\subsection{Iterative Algorithm for Solving Problem \eqref{eq:JFCS2_2}}

We are now in a position to tackle the nonconvexity of \eqref{eq:problemad} and \eqref{eq:12} by the IA framework.

\textit{Convexity of constraint \eqref{eq:problemad}:}
We first relax \eqref{eq:problemad} by considering its simple form: $\qquad p_k^i[t] \leq e^{\mathtt{eff}}_k/(1-\alpha_k[t])\tau$.
From (\ref{eq1}) and (\ref{eq9}), we can rewrite it as:
\begin{IEEEeqnarray}{cl}\label{eq13}
    p_k^i[t] & \leq \frac{E_{k}[t]}{\left(1-\alpha_k[t]\right)\tau} + \eta\|\boldsymbol{g}_k[t]\|^2\frac{\alpha_k[t] p_k^e[t]}{1-\alpha_k[t]}  \nonumber\\
    & = \frac{E_{k}[t]}{\tau}\frac{1}{1-\alpha_k[t]} + \eta\|\boldsymbol{g}_k[t]\|^2\frac{1}{4}\Big[\frac{(\alpha_k[t] + p_k^e[t])^2}{1-\alpha_k[t]} \nonumber\\
    &\quad - \frac{(\alpha_k[t] - p_k^e[t])^2}{1-\alpha_k[t]},\,\forall k\Big]\end{IEEEeqnarray}
where the second equation is due to the fact that $xy = \frac{1}{4}[(x+y)^2 - (x-y)^2].$ From \cite{NguyenTCOM17,Beck:JGO:10}, the first-order Taylor approximations of $f(x)\triangleq 1/x$ and $g(x,y)\triangleq x^2/y$  can be found as
\begin{align}
{1}/{x} &\geq f(\bar{x}) + \nabla f(\bar{x})(x-\bar{x})={2}/{\bar{x}}  - {x}/{\bar{x}^2},\label{eq:15} \\
{x^2}/{y} &\geq g(\bar{x},\bar{y}) + \big<\nabla g(\bar{x},\bar{y}), (x,y) - (\bar{x},\bar{y})\big> \nonumber\\
&= 2 \frac{\bar{x}x}{\bar{y}}  - \frac{\bar{x}^2}{\bar{y}^2}y,\, \forall x, \bar{x}, y, \bar{y}\in\mathbb{R}_+.\label{eq:16}
\end{align}
The equalities in \eqref{eq:15} and \eqref{eq:16} hold true if $x \equiv\bar{x}$ and $y \equiv\bar{y}$. We use \eqref{eq:15} and \eqref{eq:16} to iteratively replace \eqref{eq13} by the following approximate convex constraint at iteration $\kappa$:
\begin{align} \label{eq21}
    &p_k^i[t] + \eta\|\boldsymbol{g}_k[t]\|^2 \frac{(\alpha_k[t] - p_k^e[t])^2}{4(1-\alpha_k[t])} \leq \frac{E_k[t]}{\tau}\times \nonumber\\
    &\Big[\frac{2}{1-\alpha^{(\kappa)}_k}-\frac{1-\alpha_k[t]}{(1-\alpha^{(\kappa)}_k)^2}\Big] + \frac{1}{4} \Big[\frac{2(\alpha^{(\kappa)}_k+p^{e,(\kappa)}_k[t])}{1-\alpha^{(\kappa)}_k}\times\nonumber\\
    &(\alpha_k[t]+p^e_k[t])-\frac{(\alpha^{(\kappa)}_k+p^{e,(\kappa)}_k[t])^2}{(1-\alpha^{(\kappa)}_k)^2}(1-\alpha_k[t])\Big].
\end{align}

\textit{Convexity of constraint \eqref{eq:12}:}
From \eqref{eq:DATAqueue}, we can equivalently express \eqref{eq:12} as
\begin{equation} \label{eq26}
    \frac{q_k[t] + a_k[t]\tau - \lambda_k[t]}{1 - \alpha_k[t]} \leq r_k(p_k^i[t])\tau,\,\forall k.
\end{equation}
We introduce slack variables $\hat{\boldsymbol{\alpha}}[t] = \{\hat{\alpha}_k[t]\}_{\forall k}$, satisfying
\begin{equation}\label{eq:19}
    \frac{1}{1-\alpha_k[t]} \leq \hat{\alpha}_k[t],\,\forall k
\end{equation}
to convert \eqref{eq26} into the following form:
\begin{equation} \label{eq29}
    (q_k[t] + a_k[t]\tau - \lambda_k[t]) \hat{\alpha}_k[t] \leq r_k(p_k^i[t])\tau,\,\forall k.
\end{equation}
It is noted that the right-hand side of \eqref{eq29} is a concave function. For simplicity, let us introduce new variables $\boldsymbol{\psi}[t] = \{\psi_k[t]\}_{\forall k}$, satisfying
\begin{equation}\label{eq21a}
\psi_k[t]\geq q_k[t] + a_k[t]\tau - \lambda_k[t],\,\forall k
\end{equation}
which is a linear constraint. As a result, constraint \eqref{eq29} becomes $\psi_k[t] \hat{\alpha}_k[t] \leq r_k(p_k^i[t])\tau$. The function $\psi_k[t] \hat{\alpha}_k[t]$ is neither convex nor concave, however, its concave upper bound can be found as 
\begin{equation}
    \psi_k[t] \hat{\alpha}_k[t] \leq \frac{1}{2} \frac{\psi_k^{(\kappa)}[t]}{\hat{\alpha}^{(\kappa)}_k} \hat{\alpha}_k^2[t] + \frac{1}{2} \frac{\hat{\alpha}^{(\kappa)}_k}{\psi_k^{(\kappa)}[t]} \psi_k^2[t],\,\forall k
    \label{eq31a}
\end{equation}which is a convex function with respect to $\psi_k[t]$ and $\hat{\alpha}_k[t]$.   At iteration $\kappa$, \eqref{eq29} is  replaced by
\begin{equation} \label{eq31}
    \frac{1}{2} \frac{\psi_k^{(\kappa)}[t]}{\hat{\alpha}^{(\kappa)}_k} \hat{\alpha}_k^2[t] + \frac{1}{2} \frac{\hat{\alpha}^{(\kappa)}_k}{\psi_k^{(\kappa)}[t]} \psi_k^2[t] \leq r_k(p_k^i[t])\tau,\, \forall k.
\end{equation}

Next, constraint \eqref{eq:19} can be re-expressed as
\begin{IEEEeqnarray}{cl}\label{eq28}
    &1 \leq \frac{1}{4}(\hat{\alpha}_k[t] + 1 - \alpha_k[t])^2 - \frac{1}{4}(\hat{\alpha}_k[t] - 1 + \alpha_k[t])^2 \nonumber\\
    \Leftrightarrow & \sqrt{1 + \Big(\frac{\hat{\alpha}_k[t] - 1 + \alpha_k[t]}{2}\Big)^2} \leq \frac{\hat{\alpha}_k[t] + 1 -\alpha_k[t]}{2}\qquad
\end{IEEEeqnarray}
which can be cast to the second-order cone (SOC) constraint, \textit{i.e.} $\|1, \frac{1}{2}(\hat{\alpha}_k[t] - 1 + \alpha_k[t])\|_2 \leq \frac{1}{2}(\hat{\alpha}_k[t] + 1 - \alpha_k[t])$.

From the above developments, we solve the following convex program at iteration $\kappa$ of problem \eqref{eq:JFCS2_2}:
\begin{subequations} \label{eq:JFCS3Co}
    \begin{IEEEeqnarray}{cl}
	\underset{\boldsymbol{p}, \boldsymbol{\alpha}, \boldsymbol{\lambda}, \boldsymbol{\psi}, \hat{\boldsymbol{\alpha}}}{\mathrm{max}} \,& \beta \sum_{k\in\mathcal{K}}r_k(p_k^i[t]) - \Delta L^{\mathtt{UB}}[t]\label{eq:JFCS3Co_a} \\
      \st&  \ \eqref{eq:problemac},  \eqref{eq:problemaf}, \eqref{eq21}, \eqref{eq21a}, \eqref{eq31}, \eqref{eq28}.\label{eq:JFCS3Co_b} 
    \end{IEEEeqnarray}
\end{subequations}
The proposed iterative algorithm for solving  \eqref{eq:JFCS2_2} is outlined in Algorithm \ref{alg2}. In Step 2 of iteration $\kappa$, we solve \eqref{eq:JFCS3Co} by convex solvers (\textit{i.e.} CVX) to obtain the optimal solution, which is then used as the feasible point for iteration $\kappa+1$.

\begin{algorithm}[t]
\begin{algorithmic}[1]
\fontsize{9}{9}\selectfont
\protect\caption{Proposed Iterative Algorithm for Solving \eqref{eq:JFCS2_2}}
\label{alg2}
\global\long\def\algorithmicrequire{\textbf{Initialization:}}
\REQUIRE  Set $\kappa:=1$ and  generate the initial feasible points $\{\boldsymbol{p}^{(0)}, \boldsymbol{\alpha}^{(0)}, \boldsymbol{\lambda}^{(0)}, \boldsymbol{\psi}^{(0)}, \hat{\boldsymbol{\alpha}}^{(0)}\}$ for constraints in \eqref{eq:JFCS2_2}.
\REPEAT
\STATE Solve  \eqref{eq:JFCS3Co} to obtain the optimal solution $\{\boldsymbol{p}^{(*)}, \boldsymbol{\alpha}^{(*)}, \boldsymbol{\lambda}^{(*)}, \boldsymbol{\psi}^{(*)}, \hat{\boldsymbol{\alpha}}^{(*)}\}$;

\STATE Update $\{\boldsymbol{\alpha}^{(\kappa)}, \boldsymbol{\psi}^{(\kappa)}, \hat{\boldsymbol{\alpha}}^{(\kappa)}\} := \{\boldsymbol{\alpha}^{(*)}, \boldsymbol{\psi}^{(*)}, \hat{\boldsymbol{\alpha}}^{(*)}\}$,\, $p_k^{e,(\kappa)}[t] := p_k^{e,(*)}[t]$ and $p_k^{i,(\kappa)}[t] := \min\Big\{\frac{e^{\mathtt{eff},(*)}_k}{(1-\alpha_k^{(*)}[t])\tau}, \bar{P}_k\Big\},\, \forall  k$;
\STATE Set $\kappa:=\kappa+1;$
\UNTIL Convergence;\\
\STATE{\textbf{Output:}} $\{\boldsymbol{p}^{(*)}, \boldsymbol{\alpha}^{(*)}, \boldsymbol{\lambda}^{(*)}, \boldsymbol{\psi}^{(*)}, \hat{\boldsymbol{\alpha}}^{(*)}\}$.
\end{algorithmic} \end{algorithm}

\textit{Convergence and Complexity Analysis:} Based on the IA principles \cite{Beck:JGO:10}, we can show that Algorithm \ref{alg2} produces a sequence of better solutions for the approximated optimization variables, which converge to at least a local optimal solution of problem \eqref{eq:JFCS2_2}, and hence \eqref{eq:JFCS2} due to \eqref{eq:12}. For an appropriate scaling factor $\beta$, problems \eqref{eq:JFCS2} and \eqref{eq:JFCS1} share the same optimal solution \cite{neely2010stochastic} at each time-frame $t$. The computation complexity of 
Algorithm \ref{Alg1} mainly comes from Step 2 (\textit{i.e.} Algorithm \ref{alg2}). Problem \eqref{eq:JFCS3Co} involves  $6K$ scalar decision variables and $5K+1$ linear and SOC constraints. Consequently, the worst-case of per-iteration complexity of Algorithm \ref{alg2} is $\mathcal{O}\bigl(\sqrt{5K}(6K)^3 \bigr)$  \cite[Chapter 6]{Ben:2001}.

\section{Numerical Results}
\subsection{Network Setup}
We consider a network consisting of four UEs randomly distributed within a circle of $250$-m radius. PB and AP are positioned at coordinates ($-250, 0$) and ($0, 0$), respectively. Both PB and AP are equipped with $M = N = 16$ antennas. All the wireless channels follow an exponential Rayleigh distribution with the path-loss exponent of $3$. The PB has a total power budget of $P_{\max} = 43$ dBm, and the energy conversion efficiency is set to $\eta = 0.6$. Each time-frame has a $\tau = 1$ ms duration, and the system operates with a bandwidth of $W =10$ MHz. The data arrival rate for UE $k$ at time-frame $t$ is uniformly distributed in $[40,\,60]$ Mbps. Additionally, we set $E_k^{\max}=3\times 10^3$ J, $\bar{P}_k = 10$ dBm and $\underline{\gamma}^{\min} = -10$ dB.

For performance comparison, we consider the following three benchmark schemes.
\begin{itemize}
    \item Equal-power $p^e$: The power budget of PB is equally allocated to all UEs in the WET phase, \textit{i.e.} $p^e \equiv p^e_k[t] = P_{\max}/K$, $\forall t, k$.
    \item Max-power: The power harvested during the WET phase is fully utilized for information transmission in the WIT phase, \textit{i.e.} $p^i_k[t] = \min\big\{\frac{e_k[t]}{(1-\alpha_k[t])\tau}, \bar{P}_k\big\},\,\forall t, k$. 
    \item Equal-time: The time fraction allocated to the WET and WIT phases is equally divided,  \textit{i.e.} $\alpha_k[t] = 0.5,\, \forall k,t$.
\end{itemize}
The corresponding problems can be efficiently solved by Algorithm \ref{alg2}. The results are averaged over 1000 channel realizations.
\subsection{Results and Discussion}

 \begin{figure}[t]
    \centering
    \includegraphics[scale = 0.5]{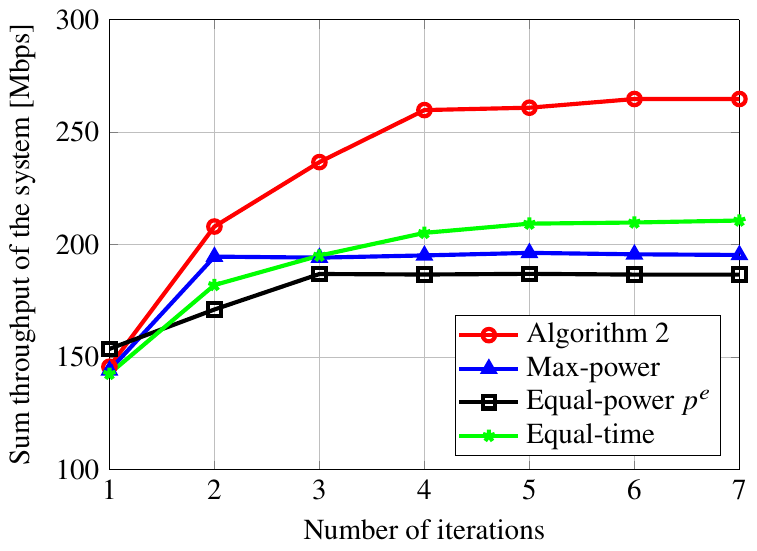}
    \caption{\small Convergence behavior of different schemes.}
    \label{fig1_Convergence}
\end{figure}
\begin{figure}[t]
    \centering
    \includegraphics[scale = 0.5]{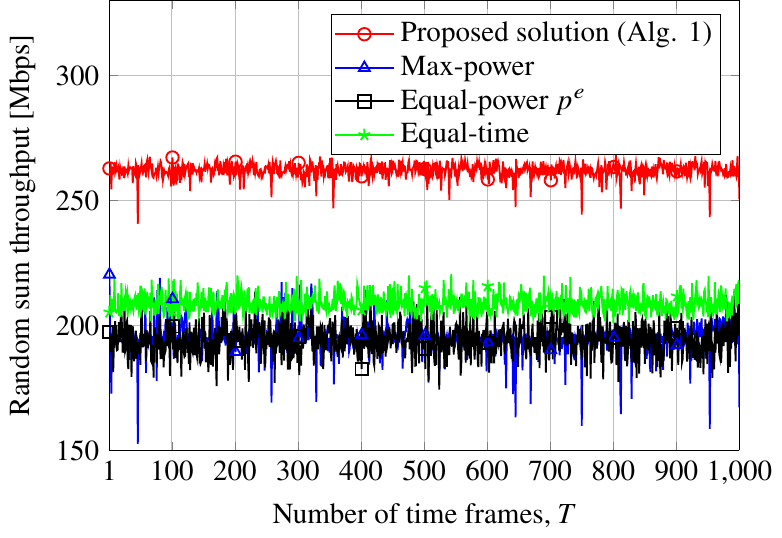}
    \caption{\small Sum throughput of difference schemes versus the number of time-frames}
    \label{fig2_RandomSumthroughput} 
\end{figure}
Fig. \ref{fig1_Convergence} illustrates the typical convergence behavior of Algorithm \ref{alg2} compared to benchmark schemes over a random channel. All considered schemes converge within about six iterations. As expected, the sum throughput is non-decreasing with each iteration. Algorithm \ref{alg2} outperforms the benchmarks, demonstrating superior performance due to the joint optimization of power allocation and time fraction. Additionally, the ``Equal-time'' scheme shows better performance than the others, confirming the critical role of power allocation in both the WET and WIT phases.

In Fig. \ref{fig2_RandomSumthroughput}, we present the sum throughput over 1000 time frames for different resource allocation schemes. As can be seen, Algorithm \ref{Alg1} significantly outperforms the benchmark schemes. Furthermore, the performance of Algorithm \ref{Alg1} remains more consistently stable around the average, demonstrating the effectiveness of the energy-aware resource allocation approach under the accumulate-then-transmit protocol in managing the uncertainties of wireless channels.

\begin{figure}[t]
    \centering
    \includegraphics[scale = 0.5]{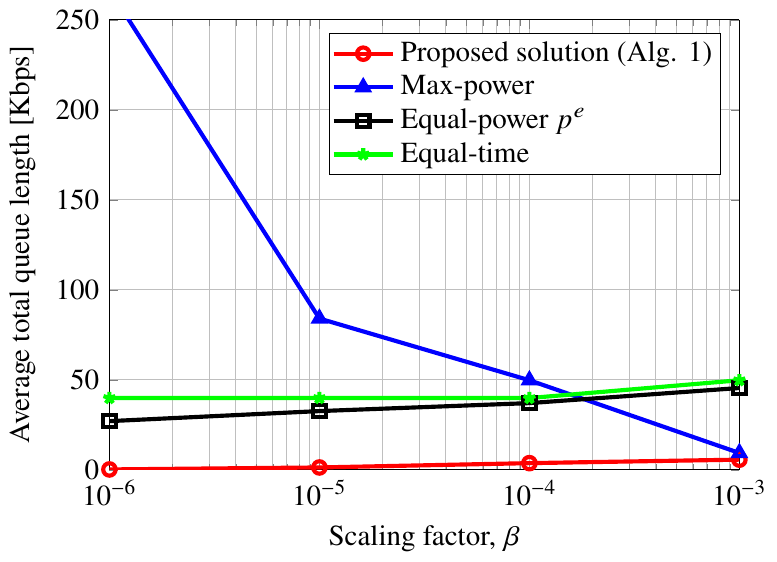}
    \caption{\small Average total queue length with respect to the scaling factor, $\beta$.}
    \label{fig3_Queuelength}
\end{figure}

\begin{figure}[!ht]%
\centering
\subfigure[Average sum of harvested power]{%
\label{fig:Convergence-a}%
\includegraphics[width=0.45\columnwidth]{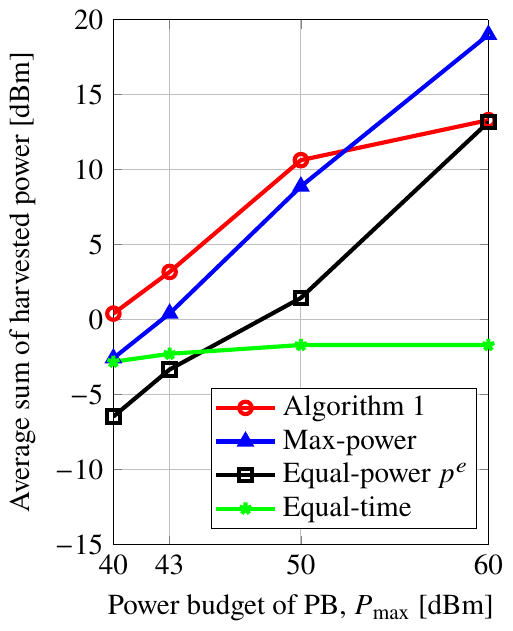}
}%
\subfigure[Average sum throughput]{%
\label{fig:Convergence-b}%
\includegraphics[width=0.45\columnwidth]{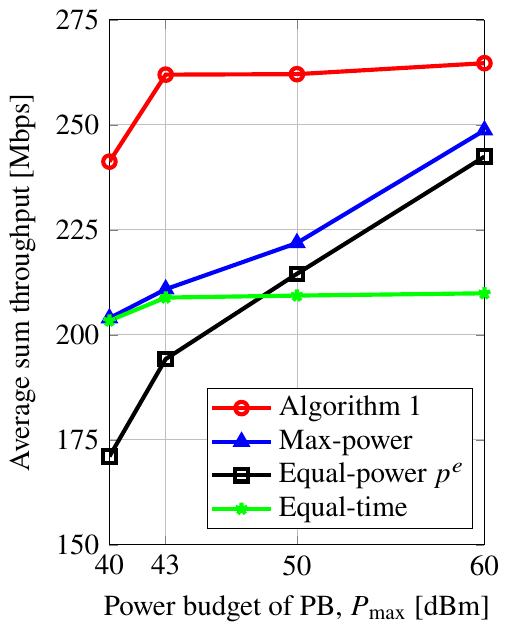}} 
\caption[]{\small Average sum of harvested power and throughput versus the total power budget at PB, $P_{\max}$.}%
\label{fig4_AverageSumPowerandThroughput}%
\end{figure}

Fig. \ref{fig3_Queuelength} shows the impact of the scaling factor, $\beta$, on the average total queue length. We vary $\beta$ within the range of $[10^{-6},\,10^{-3}]$. For the ``Max-power" scheme, the total queue length decreases as $\beta$ increases, as this scheme prioritizes maximizing throughput in each time-frame. In contrast, the total queue length for the other schemes increases with $\beta$. This is because a larger $\beta$ gives more weight to the network utility function $U(\boldsymbol{p}|\boldsymbol{s})$, resulting in a higher Lyapunov drift. Once again, the proposed Algorithm \ref{Alg1} outperforms the other schemes. In our simulations, we select $\beta = 10^{-4}$, which provides the best balance of fairness between the schemes.

Finally, in Fig. \ref{fig4_AverageSumPowerandThroughput}, we plot the average sum of harvested power and throughput as a function of $P_{\max}\in[40,\, 60]$ dBm. As anticipated, a higher power budget improves performance for all schemes. Notably, Algorithm \ref{Alg1} outperforms the benchmark schemes in terms of harvested power in Fig. \ref{fig4_AverageSumPowerandThroughput}(a), particularly in the lower-to-mid power budget range, due to its efficient optimization of power and harvesting time allocation during the WET phase. In Fig. \ref{fig4_AverageSumPowerandThroughput}(b), Algorithm \ref{Alg1} quickly reaches saturation when $P_{\max} > 43$ dBm, offering a more efficient trade-off between data and energy queues.

\section{Conclusion}
This paper addressed the energy-aware resource allocation problem in EH-powered wireless sensor networks. We present the novel optimization of the accumulate-then-transmit protocol that jointly optimizes power allocation and energy harvesting time. Utilizing IA and NUM techniques, the proposed iterative optimization algorithm ensures convergence to a local optimum while maximizing long-term system throughput and effectively managing dynamic data and energy queues. Numerical simulations are provided to demonstrate the effectiveness of the proposed approach in terms of the total sum throughout and queue length. Future work will consider a non-linear EH model for a more practical representation.

\bibliographystyle{IEEEtran}
\bibliography{Bibliography}

\end{document}